\documentclass[useAMS,usenatbib,aas_macros]{mn2e}

\usepackage{hhline}
\usepackage{lscape}
\usepackage{graphicx}
\usepackage{colortbl}
\usepackage{amssymb}
\usepackage{amsmath}
\usepackage{natbib}
\usepackage{times}
\usepackage{aas_macros}
\usepackage{threeparttable}

\newcommand{\msun}{~\mathrm{M}_{\odot}}

\newcommand{\J}{J$_{21}$~}
\usepackage{color}
\defcitealias{Agarwal12}{A12}
\defcitealias{Dijkstra2014a}{D14}
\defcitealias{habouzit16}{H16}
\defcitealias{Sesana:2007p2147}{S07}

\def\simpropto{\lower.2ex\hbox{$\; \buildrel \propto \over \sim \;$}}
\def\ltsim{\lower.5ex\hbox{$\; \buildrel < \over \sim \;$}}
\def\gtsim{\lower.5ex\hbox{$\; \buildrel > \over \sim \;$}}

\def\etal{{\it et al.}~}
\begin{document}
\title[GW from DCBHs]{Gravitational Wave Signals from the First Massive Black Hole Seeds}
\author[T. Hartwig \etal]{ Tilman Hartwig$^{1,2}$\thanks{E-mail: Tilman.Hartwig@ipmu.jp},
  Bhaskar Agarwal$^{3}$   \& John A. Regan$^{4}$\thanks{Marie Sk\l odowska-Curie Fellow} \\ \\
  $^1$Department of Physics, School of Science, University of Tokyo, Bunkyo, Tokyo 113-0033, Japan\\
$^2$Kavli IPMU (WPI), The University of Tokyo, Kashiwa, Chiba 277-8583, Japan\\
$^3$Universit{\"a}t Heidelberg, Zentrum fur Astronomie, Institut fur Theoretische Astrophysik, Albert-Ueberle-Str. 2, D-69120 Heidelberg\\
  $^4$Centre for Astrophysics \& Relativity, School of Mathematical Sciences, Dublin City University, Glasnevin, Ireland
  \\}

% ----------------------------------------------------------------

\date{\today}
\pagerange{\pageref{firstpage}--\pageref{lastpage}} \pubyear{0000}
\maketitle

\label{firstpage}

% ----------------------------------------------------------------
\begin{abstract}
  Recent numerical simulations reveal that the isothermal collapse of pristine gas in atomic cooling haloes
  may result in stellar binaries of supermassive stars with $M_*\gtrsim 10^4\msun$. For the first time,
  we compute the in-situ merger rate for such massive black hole remnants by combining their abundance
  and multiplicity estimates. For black holes with initial masses in the range
  $10^{4-6} \msun$ merging at redshifts $z \gtrsim 15$ our optimistic model predicts that LISA should be
  able to detect 0.6 mergers per year. This rate of detection can be attributed, without confusion,
  to the in-situ mergers of seeds from the collapse of very massive stars. 
  Equally, in the case where LISA observes no mergers from heavy seeds at $z \gtrsim 15$ we can constrain
  the combined number density, multiplicity, and coalescence times of these high-redshift systems. This
  letter proposes gravitational wave signatures as a means to constrain theoretical models and processes
  that govern the abundance of massive black hole seeds in the early Universe.
 \end{abstract}
% ----------------------------------------------------------------


%%%%%%%%%%%%%%%%%%%%%%%%%%%%%%%%%%%%%%%%%%%%%%%%%%%%%%%%%%%%%%%%%%%%%%%%%%
%%%%%%%%%%%%%%%%%%%%%%%%%%%%%%%%%%%%%%%%%%%%%%%%%%%%%%%%%%%%%%%%%%%%%%%%%%

%%%%%%%%%%%%%%%%%%%%%%%%%%%%%%%%%%%%%%%%%%
\section{Introduction}
%%%%%%%%%%%%%%%%%%%%%%%%%%%%%%%%%%%%%%%%%%
%(tbd: unify acronyms)
The two aLIGO detectors \citep{harry2010} observed the first detection of a black hole binary (BHB)
on September 14, 2015. This was an unprecedented event and heralded the dawn of gravitational wave
(GW) astronomy. The event, GW150914, involved two black holes with masses $36^{+5}_{-4} \msun$ and
$29^{+4}_{-4} \msun$ coalescing at z = $0.09^{+0.03}_{-0.04}$ \citep{GW150914}.
While the sensitivity of
aLIGO is limited to the detection of stellar mass sized BHBs, future gravitational wave detectors
have the potential to detect more massive systems out to much larger distances and redshifts. LISA
\citep{eLISA} due for launch in 2034 will be able to detect BHB mergers out to z $\gtrsim 20$ opening
a window of observation on black hole growth and evolution in the early Universe. \\
\indent The observation of supermassive black holes (SMBHs) shining as quasars at redshifts
greater than 6 \citep{Fan:2003p40, Mortlock:2011p447, Venemans:2013p3633,2015Natur.518..512W,Banados_2018}
has led to a theoretical challenge in astrophysics. How could such massive objects form so early in the
Universe? Stellar mass black hole seeds, similar in mass to those detected by GW150914, are
expected to form as the remnants of the very first generation of stars \citep{Heger:2003p23},
however, these seed black holes are also expected to be born starving with little prospect of
growing substantially in the early Universe \citep{Johnson:2007p48,
  Milosavljevic:2009p779, Alvarez:2009p778,smith18}. Alternatively, a heavy seed model has been
proposed that potentially overcomes this early bottleneck. Massive black hole seeds born
from the remnants of
(super-)massive stars with masses $M_* \gtrsim 10^4 \msun$ in atomic cooling haloes (ACHs) have the potential
to grow at higher accretion rates (at least initially) compared to lower mass stellar mass seeds. Theses heavy seeds have
been dubbed ``Direct Collapse Black Holes'' (DCBHs). The deeper potential wells within which they are
born are expected to provide the continuous supply of matter required for the black hole to achieve
masses close to a billion solar masses by a redshift of six. 
 Determining the actual number densities of heavy seeds is currently an area of intense research \citep[][hereafter \citetalias{Agarwal12}, \citetalias{Dijkstra2014a}, \citetalias{habouzit16}]{Agarwal12,Dijkstra2014a,habouzit16} and determining what
fraction of massive black holes originate from heavy seeds an outstanding problem. Heavy seeds
may therefore be the progenitors of all massive black holes, or a small sub-population. \\
\indent Over the past decade, although increasingly sophisticated simulations and semi-analytic modelling has been employed in an attempt to understand the seed formation mechanisms for SMBHs, the task remains challenging due to lack of observational data.
Recent progress on the modelling
of supermassive stars (SMSs) suggests that a high accretion rate onto a protostellar core does lead to the formation of an SMS \citep{Hosokawa:2008p20, Hosokawa:2013p3513,
  Woods_2017,haemmerle18}, which would be the ideal progenitor for a heavy seed. 
 In a cosmological context, studies
(\citetalias{Agarwal12}, \citealt{Agarwal14}, \citetalias{Dijkstra2014a}, \citetalias{habouzit16}, \citealt{chon18})
% \citep{Agarwal12,Agarwal14,Dijkstra2014a,habouzit16,chon18}
 have shown that in rare 
regions of the Universe, where ACHs
can remain metal free and the formation of molecular hydrogen is suppressed (due to a strong photodissociating background for example), conditions become
conducive to the formation of an SMS. Moreover, SMS forming regions appear to favour the formation of
a small number of very massive fragments \citep{chon18, Regan_2018a, Regan_2018b} all, or at
least some of which, can potentially form massive seeds.
If the SMSs in these multiple systems form as tight binaries that do not merge until after the
stars have collapsed into black holes, then the resulting black hole seeds will have initial masses
greater than $10^4 \msun$, candidates ideal for detection by LISA. \\
\indent The goal of this paper is to generate templates of DCBH in-situ mergers that can
subsequently be compared against LISA detection rates to constrain both the formation scenario and
abundance of DCBHs. To achieve this we model the expected
number density of DCBH formation sites to constrain the total number of DCBHs expected per unit
redshift. We combine this with an estimate of the fragmentation rate within DCBH haloes to obtain the distribution of BHBs and their coalescence times within haloes
that host multiple DCBH formation sites. We 
focus on mergers of heavy seeds ($10^4 \msun < M_{BH} < 10^6 \msun$)
within the halo in which they are born. We do not consider
mergers of black holes as a consequence of galaxy mergers. From the expected number density of
DCBH formation haloes  and the fragmentation probabilities within these haloes
we can quantify the expected rate of DCBH mergers that will
be detectable by LISA due to SMS multiplicity.

%%%%%%%%%%%%%%%%%%%%%%%%%%%%%%%%%%%%%%%%%%
%\section{Methodology}
%%%%%%%%%%%%%%%%%%%%%%%%%%%%%%%%%%%%%%%%%%
\section{Number Density Distribution of DCBHs} \label{NumberDensity}

In order to obtain the abundance of DCBH seeds, we use the results of A12, D14 and H16. We focus on two scenarios with
J$_c = 30$ J$_{21}$\footnote{\J is the LW flux in units of $10^{-21}$ erg cm$^{-2}$
  s$^{-1}$ Hz$^{-1}$ sr$^{-1}$.} and 300 \J respectively, where J$_c$ is the critical Lyman-Werner (LW) flux
required to induce DCBH formation. The exact value of J$_c$
that can facilitate DCBH formation in an ACH is still unknown. What complicates the
matter further is that a single value of J$_c$ 
is not representative of the chemo-thermodynamical processes
that lead to the formation of a DCBH, namely the photo-destruction of H$^-$ and H$_2$
\citep{WolcottGreen:2012p3854,Sugimura:2014p3946, Agarwal15a,2016MNRAS.tmp..708A}. Therefore, we use abundance
estimates from the literature and employ two values of J$_c$ as our extreme limits. In doing
so we are implicitly assuming that other process may either augment or replace the effect of an
external LW field to produce similar results (e.g. streaming velocities, see \citealt{Tanaka_2014, Hirano_2017, Schauer_2017}, or rapid accretion, see \citealt{Yoshida:2003p51}). Thus, J$_c = 30$ \J can be viewed as
a scenario in which DCBHs are relatively common and would result in a sufficient number of massive
seeds for the entire MBH population. On the other hand, J$_c = 300$ \J would result in a number
density of heavy seeds that could only seed a sub-population of massive black holes (perhaps the
highest redshift quasars) and can thus be viewed as a scenario
where large initial mass black holes are rare. 
We now describe our steps going from the theoretical estimates of abundance of DCBH seeds to the observed number of events.

%\begin{itemize}
%\item[]
Using the models and predictions from the redshift distributions of the newly formed
  DCBH seeds per unit comoving Mpc$^3$ found in A12, D14, H16 we calculate a `rate'
  of formation at time step `i' in units of cMpc$^{-3}$ 
\begin{equation} \label{pertimepervol}
\left. \frac{\mathrm{d}n}{\mathrm{d}t\mathrm{d}V}\right|_i = \frac{N_\mathrm{new,i}}{t_{i} - t_{i-1}} \ \rm cMpc^{-3} ,
\end{equation}
where N$_{new}$ is the number of newly formed DCBH sites identified in the simulations for a given value of J$_c$, and $\mathrm{d}V$ denotes the co-moving volume normalisation. 
%\jr{Bhaskar - what is the typical $\Delta t$ used here?}
%To get the rate of events per unit time per unit redshift from any given redshift in the observer's frame of reference (i.e. as seen at z=0), we  follow \citealt{magg16} (eq. 8)
To get the DCBH formation rate per unit time per unit redshift from any given redshift in the observer's frame of reference (i.e. as seen at z=0), we  follow \citealt{magg16} (eq. 8)
\begin{equation} \label{eqn:mergerrate}
\frac{\mathrm{d}n}{\mathrm{d}z\mathrm{d}t} = \frac{4\pi}{1+z}R_z^2 \frac{\mathrm{d}R_z}{\mathrm{d}z}\frac{\mathrm{d}n}{\mathrm{d}t\mathrm{d}V},
\end{equation}
where R$_z$ is the comoving distance at a given redshift.
%\end{itemize}

%In the top panel of Figure \ref{fig:ndens} we plot the comoving number density extracted from the literature.
%Both \citetalias{Dijkstra2014a} and \citetalias{habouzit16} provide estimates for the number densities based on a J$_c$ value of 300 while \citetalias{Agarwal12}, \citetalias{Dijkstra2014a}, and \citetalias{habouzit16} provide estimates for an external LW field with J$_c$ = 30 \J (blue dotes). The pink dots are the values
%found for J$_c$ = 300 \J. Multiplying the J$_c$ = 30 \J values by $10^{-5}$ reproduces the values for
%J$_c$ = 300 \J -- this is shown using the red dots and allows us to scale any results
%between J$_c$ = 30 \J and J$_c$ = 300 \J with ease. This reflects the combined effect of the probability distribution function (PDF) of J$_c$ that 
%irradiates pristine ACHs and the abundance of metal--poor ACHs, where the local variation of J$_c$ is accounted for \citep[e.g.][A12, D14, H16]{Dijkstra:2008p45}. In the bottom panel of \ref{fig:ndens} we plot the number of DCBH seeds per unit time per
%unit volume i.e. equation \ref{pertimepervol}. The formation rate initially increases, peaks
%approximately 500 Myr after the Big Bang and then decreases as we approach reionisation coinciding
%with the complete metal enrichment of the Universe and the expected end of DCBH formation. We now move on to
%discuss the merging of in-situ DCBHs and from there the merger rates of DCBHs. \\

\begin{figure} 
\centering
\includegraphics[width=0.47\textwidth,trim=0mm 10mm 10mm 149.05mm,clip]{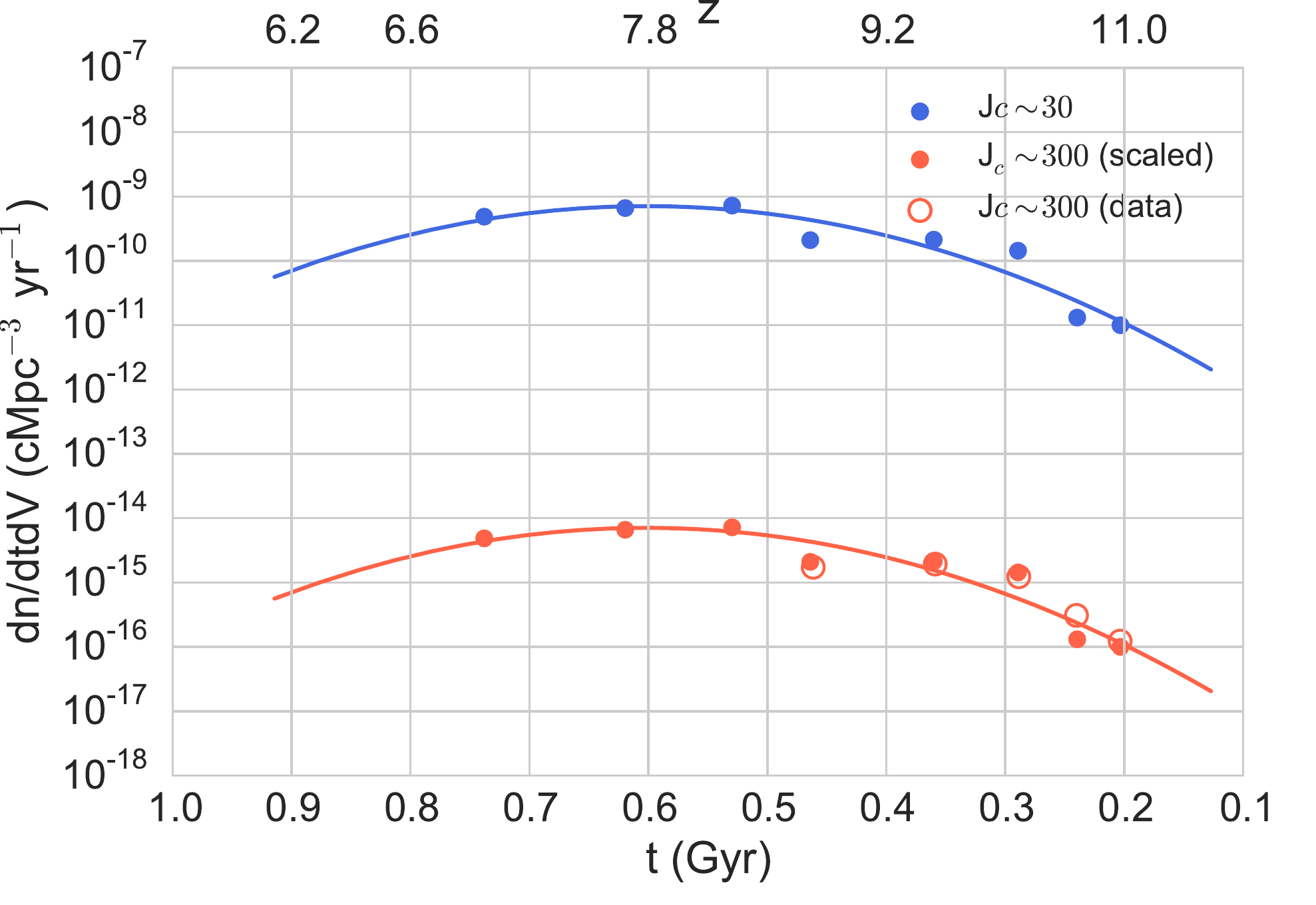}
%\caption{\textit{Top Panel:} Number density of DCBHs from various models in the literature using a J$_{c}\sim30$ in blue and J$_{c}\sim100$ \J in pink dots. We scale the data for J$_c\sim30$ \J by multiplying the data points with $10^{-5}$ (red) in order to mimic the data for J$_c\sim100$ \J. \textit{Bottom Panel:} The rate of formation of DCBHs per unit time per unit cMpc$^3$, data (circles) and fit (lines).}
\caption{The rate of formation of DCBHs per unit time per unit cMpc$^3$, data (circles) and fit (lines). %The rates employed in our calculation (blue and red lines) are derived from the estimate of number density of DCBHs found in the literature using a J$_{c}\sim30$ (blue filled circles) and J$_{c}\sim300$ \J (red open circles) respectively. 
To obtain the fit for J$_{c}\sim 300$, we scale the data for J$_c\sim30$ \J (blue filled circles) by multiplying the data points with $10^{-5}$ in order to mimic the data for J$_c\sim300$ \J  (red filled circles). Actual data for J$_c\sim300$ \J is shown in red open circles.}
\label{fig:ndens}
\end{figure}

In Figure \ref{fig:ndens} we plot the number of DCBH seeds per unit time per
unit volume i.e. equation \ref{pertimepervol}, generating a fit using a second order polynomial.
The fit is obtained from the estimate for the number density of DCBHs based on J$_c$ = 30 \J (in blue filled circles, \citetalias{Agarwal12}, \citetalias{Dijkstra2014a}, and \citetalias{habouzit16}), and can be written as
\begin{equation}
\mathrm{Log}\frac{\mathrm{d}n}{\mathrm{d}t\mathrm{d}V} = -11.26t^2  + 13.55t  -13.23
\end{equation}
where $t$ is the age of the Universe at which the rate is computed.
The number density of pristine ACHs scales approximately as J$_{21}^4$ \citep[D14,][]{Inayoshi_2015}   hence rescaling from J$_c = 30$ to J$_c = 300$
requires a normalisation factor of $10^{-5}$. This is shown using the red filled circles. The original data points are shown in red open circles.
%Multiplying the J$_c$ = 30 \J values by $10^{-5}$ approximately reproduces the values for
%J$_c$ = 300 \J -- this is shown using the red filled filled circles and allows us to scale any results
%between J$_c$ = 30 \J and J$_c$ = 300 \J with ease. 
%The fraction of haloes exposed to a given value of LW flux falls of as roughly J$_{21}^4$ with increasing \J. 
The scaling reflects the combined effect of the probability distribution function (PDF) of J$_c$ that irradiates ACHs and the abundance of such metal--poor haloes, where the local variation of J$_c$ is accounted for \citep[e.g.][A12, D14, H16]{Dijkstra:2008p45}. The formation rate initially increases, peaks
approximately 500 Myr after the Big Bang, and then decreases as we approach reionisation coinciding
with the complete metal enrichment of the Universe and the expected end of DCBH formation.
%We now move on to discuss the merging of in-situ DCBHs and from there the merger rates of DCBHs.

\section{Binary DCBHs} \label{fbin}
SMSs have been invoked as possible progenitors of massive black hole seeds
\citep{Rees_1978, Eisenstein:1995p870, Bromm:2003p22}.  Their
large initial masses combined with their likely location at the centre of high accretion flow makes
them ideal candidates for being the seeds for SMBHs. % \citep{Haiman_2006}.
While the central idea of
SMS formation is that a monolithic collapse occurs, recent high resolution simulations have shown that in fact mild fragmentation occurs during the initial collapse
\citep{Latif:2013p3629,2015MNRAS.446.2380B,latif16,chon18, Regan_2018a, Regan_2018b}. Typical separations between the massive stars that form
are between a few hundred and a few thousand AU while the typical masses of the stars that form
is between $10^4 \msun$ and $10^5 \msun$.
Simulations that can fully resolve the formation of SMSs and follow
their progress self-consistently to the formation of a black hole seed are not yet viable, let alone be performed
for a significant number of cases. Nonetheless, a small number of fragments are expected and
if the fragments are not ejected through three body interactions during the early collapse then BHBs
may indeed be commonplace in DCBH host haloes \citep{Regan_2018b}. We constrain our seed models to
black holes with masses between $10^4 \msun$ and $10^6 \msun$ which tallies with the  expected mass
range of DCBH seeds \citep{Lodato:2007p869, 2014MNRAS.443.2410F}. Meanwhile, light seeds born from the
remnants of Pop~III seeds are not expected to grow quickly enough to achieve masses within this range
before a redshift of 9 \citep{SmithCR7}. Therefore, any black hole mergers detected
at high-z and with seed masses in the DCBH window can be attributed to
heavy mass seeds. \\
\indent After the formation of a BHB, its separation shrinks due to the loss of angular
momentum until coalescence of the two black holes occurs. We are primarily interested in BHs
with a short coalescence time of $\lesssim 100$\,Myr because later merger events that
originate from SMS binaries may be confused with events of another origin, such as the merger of
central BHs after a galaxy merger \citep[e.g.][hereafter \citetalias{Sesana:2007p2147}]{Sesana:2007p2147}. The two main
processes to extract angular momentum from a BHB are GWs and dynamical interactions. The emission of GWs as a mechanism for hardening a binary can only
become relevant at sub-AU scales \citep{peters63} and hence are not the dominant source
of angular momentum loss in this context. Rather, the typical separation of
supermassive stellar binaries that form in DC halos are of the order of $100-10000$\ AU. During the
stellar binary evolution, this separation can shrink due to angular momentum loss, especially during
a common envelope phase \citep{belczynski17}. Unfortunately, we do not yet know the
distribution of initial binary separations, the exact stellar binary evolution, or the PDF of coalescence times for SMS remnant BHs. However,
the coalescence time PDF for stellar binaries at zero metallicity peaks at short times ($<100$\,Myr,
\citealt{hartwig16,belczynski17,inayoshi17}) or is logarithmically flat
\citep{kinugawa14,kinugawa16}. This indicates that the majority of binary systems merge
shortly after their formation. Also dynamical interactions with stars of a central stellar
cluster \citep{kashiyama16,hirano18} {or triple interactions in a small cluster of BHs \citep{bonetti18,ryu18}} can harden the BHB and consequently shorten the time until coalescence. For a typical DCBH host halo, \citet{chon18} estimated the
characteristic time for stars to remove the angular momentum from the BHB system via
scattering to be of the order of $10$\,Myr.\\
\indent To simplify our calculation we introduce an efficiency parameter $f_\mathrm{bin}$.
It subsumes all of the internal physical parameters of
the BHB formation and merger process that are currently inaccessible to both
observation and theory. In our fiducial case we make the strong assumption that all DCBH systems
form as binaries and that the binaries merge instantaneously and therefore we set
$f_\mathrm{bin} = 1$. In cases where the binary fraction is not equal to unity or the merger of the
binary DCBH system is long compared to the Hubble time, $f_\mathrm{bin}$
must be lowered accordingly. Putting more quantitative estimates on the PDF for the extraction of
angular momentum and for the mutiplicity of heavy seed formation systems is not yet possible and
must await more detailed simulations of the formation and
evolution of heavy seed models.

\begin{figure}
\centering
\includegraphics[width=0.47\textwidth,trim=1mm 5mm 5mm 5mm,clip]{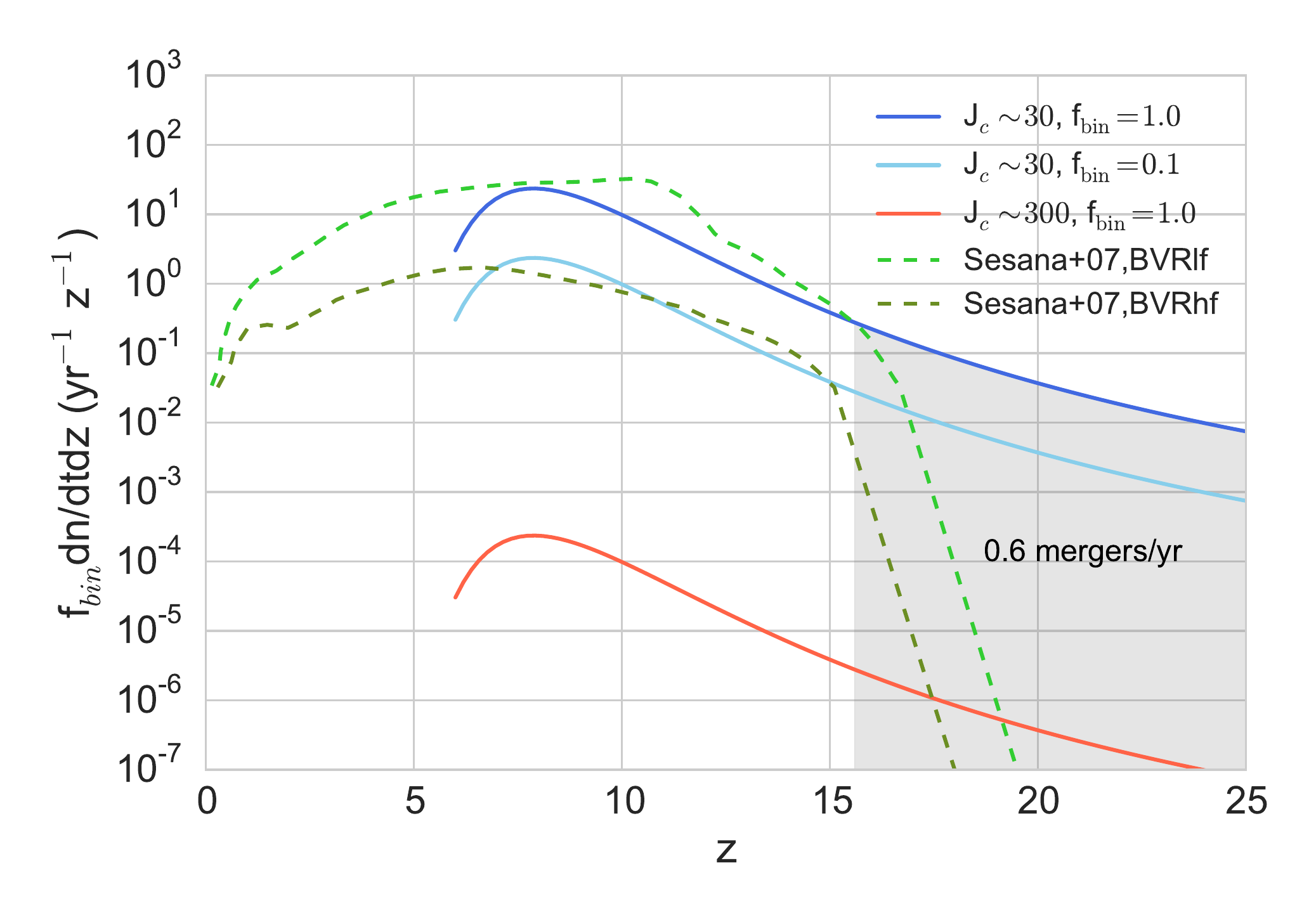}
\caption{Rates for the merging of BHB from supermassive stellar binaries as a function of redshift compared to the models by \citetalias{Sesana:2007p2147}. We show our results for different values of the critical LW flux and for different values of $f_\mathrm{bin}$, which quantifies the fraction of DCBH formation sites that host BHB that merge on a short time scale. Only our optimistic scenario with J$_\mathrm{c}\sim30$ \J and $f_\mathrm{bin}=1$ can produce a population of BHB mergers at $z \gtrsim 15$ that are clearly distinguishable from other channels of BHB formation. The total rate of such uniquely identifiable BHB mergers is $\sim 0.6$ per year, highlighted in grey.}
\label{fig:dndtdz}
\end{figure}

%%%%%%%%%%%%%%%%%%%%%%%%%%FIGURE 4%%%%%%%%%%%%%%%%%%%%%%%%%%%%%%%%%%%%%%
\begin{figure}
\centering
\includegraphics[width=0.47\textwidth,trim=0mm 5mm 5mm 5mm,clip]{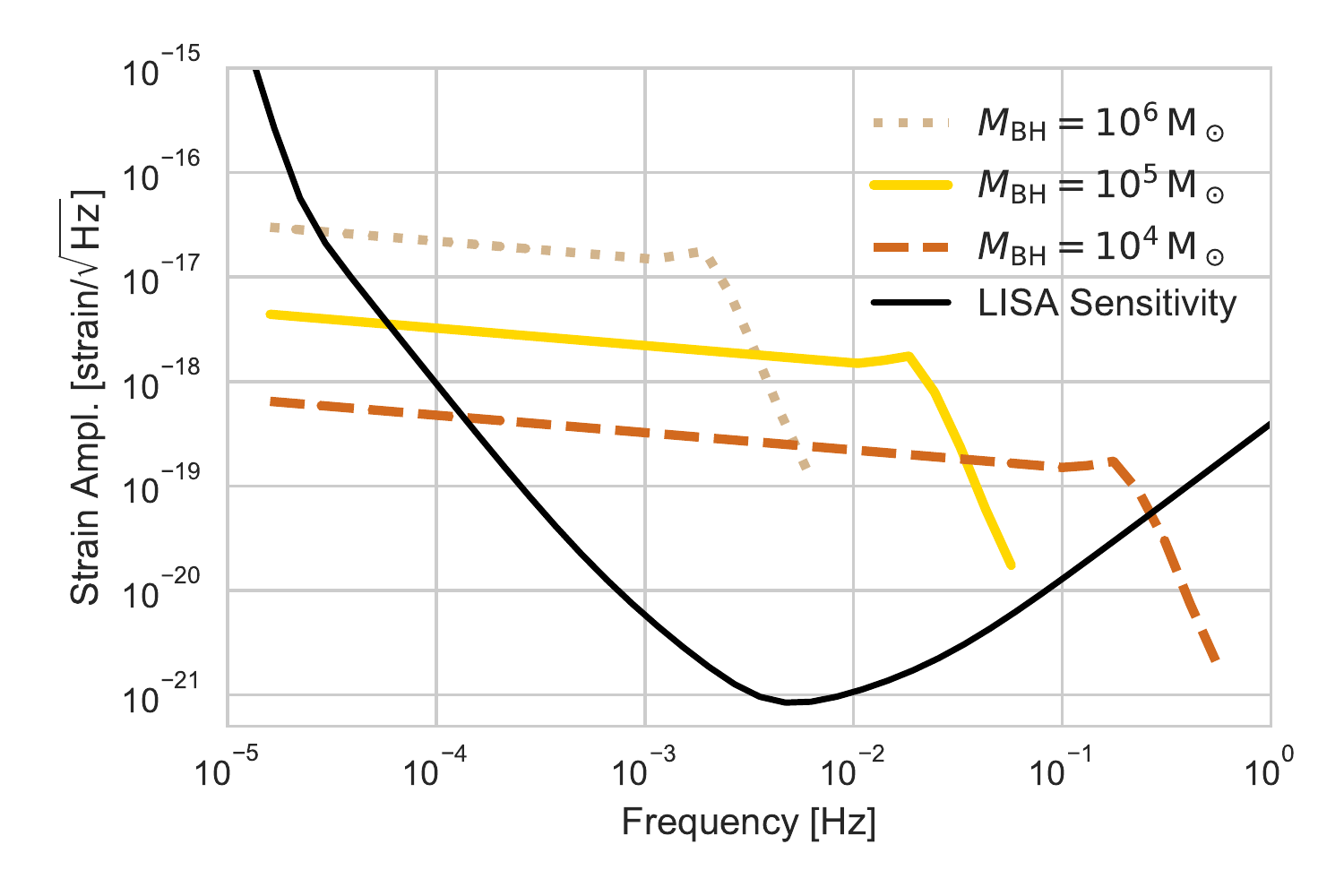}
\caption{The characteristic strain from the merger of equal mass BHBs at $z = 15$. The black line shows LISA's expected sky-averaged sensitivity, based on an arm length of $2.5$\,Mkm \citep{Babak_2017}. The other lines give the characteristic strain produced by a merger of two seed black holes, based on the waveforms for non-spinning BHBs by \citet{ajith08}, which takes into account the inspiral, merger, and ringdown stages of the coalescence. LISA will become sensitive to the merger at a frequency, f $\gtrsim 10^{-4}$\,Hz.}
%The points indicate a binary separation corresponding to the innermost stable circular orbit $R_\mathrm{isco}=6GM_\mathrm{BH}c^{-2}$. Beyond this frequency, our strain evolution should be treated with caution since we do not include post-Newtonian corrections (tbd).}
\label{fig:strain}
\end{figure}

%Comment Tilman: if we also redshift the frequency of the binary, the frequency range where binaries become visible to LISA ($10^{-4}-10^{-3}$\,Hz) is in the source frame approximately $10^{-5}-10^{-4}$\,Hz. In the most optimistic case of a $10^6\msun$ binary that enters the LISA detection limit at $10^{-4}$\,Hz in the observer rest frame is is visible for $\sim 300$\,yr. In the most pessimistic case of a $10^4\msun$ binary is becomes visible to LISA at $\sim 10^{-3}$\,Hz in the rest frame and remains visible for $\sim 10$\,yr. Generally, more massive binaries enter the LISA sensitivity curve at lower frequencies but have shorter coalescence time.

%%%%%%%%%%%%%%%%%%%%%%%%%%%%%%%%%%%%%%%%%%%%%%%%%%%%%%%%%%%%%%%%%%%%%%%%%%%
\section{Merger Rates for DCBHs} \label{MergerRates}
In Figure \ref{fig:dndtdz} we plot the merger rate of DCBHs in the range J$_c$ = 30 \J to J$_c$ = 300 J$_{21}$.
The rates for in-situ mergers of DCBHs are plotted based on equation \ref{eqn:mergerrate} for different values
of both J$_c$ and $f_\mathrm{bin}$.
The merger rate increases rapidly up to z $\sim 7$ after which it falls off as the conditions
required for DCBH formation become less favourable. Overplotted on the same
figure are the expected merger rates for massive black holes taken from \citetalias{Sesana:2007p2147} based on the models of
\citealt{BVR06} (BVR). These models are plotted as black and grey lines according to the feedback
model assumed. In \citetalias{Sesana:2007p2147} the authors, using the BVR model, consider BH mergers as a consequence of
galaxy mergers. In the BVR scenario the BHs form via runaway, global dynamical
instabilities in metal-free ACHs and gain mass via mergers and gas accretion. BVR
argue that gas-rich ACHs with efficient cooling and low angular momentum are prone to the so-called
``bars-within-bars'' mechanism \citep{Shlosman_1989}. %The effectiveness of the model relies on the amount of gas available to feed the disc and hence operates more effectively in ACHs.
Moreover, they argue that the process will naturally end when star formation becomes widespread in the
disc.
Our model examining the in-situ binary mergers is thus complementary to the BVR model that examines the merger of BHs through galaxy
mergers. \\
\indent From Figure \ref{fig:dndtdz} we expect up to $\sim 80$ mergers per year from supermassive
binary stars. In this most optimistic case (solid blue line),
with J$_c$ = 30 \J and $f_\mathrm{bin}$ = 1, the number of mergers per redshift per year are comparable to
those of BVR. The number of mergers reduces linearly with $f_\mathrm{bin}$ (light blue line). In both models the majority of mergers occur
around $z\sim 10$ and are therefore not distinguishable from each other.
Only at $z \gtrsim 15$ where our proposed channel dominates, with detection rates of $\sim 0.6$
per year, can we distinguish between the models
for in-situ mergers compared to galaxy mergers. In the BVR model detections beyond z $\sim 15$ should
be exceedingly rare due to the fact that seeds must form and their host galaxy must then merge with
another galaxy before a massive BHB merger process can begin. In our model the BHB binary is
available for merging instantaneously and hence the model extends to higher redshift. In a ten year
LISA mission our model predicts $6.1 \pm 2.5$ BHB mergers with z
$\gtrsim 15$. For J$_c$ = 300 the number of mergers expected drops by a
factor of $10^5$.\\

%%%%%%%%%%%%%%%%%%%%%%%%%%%%%%%%%%%%%%%%%%
\section{Determining the DCBH Number Density with LISA}
%\section{LISA sensitivities: Determining the DCBH Number Density from LISA}
%%%%%%%%%%%%%%%%%%%%%%%%%%%%%%%%%%%%%%%%%%
The current uncertainties surrounding the DCBH model mean that we cannot break the degeneracy between
the product of the DCBH number density (modelled here by assuming a critical LW field J$_c$) and
$f_\mathrm{bin}$. As discussed in detail above $f_\mathrm{bin}$ accounts for the fraction of DCBH systems that host
a binary that merges to produce a gravitational wave signal. A detection of BHB mergers at
$z\gtrsim 15$ could constrain the product of the DCBH number density and $f_\mathrm{bin}$. As discussed in
\S \ref{MergerRates} over the 10 year lifetime of LISA we expect $6.1 \pm 2.5$ events. Further
numerical modelling of the hosts of DCBH systems will be
required to provide greater insight into the exact value of $f_\mathrm{bin}$. Nonetheless, by employing
an optimistic value of $f_\mathrm{bin} = 1$ we can estimate the expected detection rates from LISA. 

%TBD (Enrico): Moreover, at z=15 the uncertainty on dL does not only come from the FIM, but also from weak lensing. The error on dL from lensing alone is 23\% at z=15 \citep{tamanini18}.
%Again, that needs to be summed with the FIM error (eq 7.2).%So I think telling a BH at z=10 from one at z=15 will be very difficult.

LISA is ideally suited for detecting the gravitational wave signal from black hole masses in the
range of $\sim 10^4 \leq M_\mathrm{tot}/\msun \leq 10^6\msun$ at redshifts up to z $\sim$ 20 with
signal-to-noise ratio of SNR $>10$ \citep{lisa13, Babak_2017}. Predicting the precision of the inferred luminosity distance (and hence redshift) requires numerical simulations \citep{porter15} or a detailed fisher matrix analysis \citep{klein16}, which is beyond the scope of this paper. In addition, the uncertainty on the luminosity distance, and therefore on the inferred event redshift, due to weak lensing is 23\% at $z=15$ \citep{tamanini18}.
%We have confirmed that for a comparable population of massive BHB at $z>7$, the luminosity distance can be estimate with a relative error of $<30\%$ for more than half of the detections \citep{klein16}. 
In Figure \ref{fig:strain} we plot the characteristic strain amplitude for black hole seed mergers of two equal mass black holes at $z = 15$. The LISA sensitivity is shown by the black curve using a LISA arm length of 2.5 million kilometers connected by six laser links \citep{Babak_2017}. Noise in the detector is accounted for following the analytical considerations of \cite{Babak_2017} that includes noise contributions for low-frequency noise, local interferomter noise and shot noise. The coloured lines illustrate the strain evolution based on waveforms for equal-mass, non-spinning BHBs \citep{ajith08}. We focus on equal-mass BHBs since the mass ratio $q=M_1/M_2$ with $M_1<M_2$ contributes only as a second-order correction to the coalescence time and strain amplitude compared to an equal-mass binary with the same total mass. The LISA Pathfinder mission has spectacularly demonstrated the capabilities of the hardware with obtained sensitivities below the estimates for LISA. Therefore, our sensitivity curve can be seen as a conservative limit \citep{LPF18}.\\
\indent In \S \ref{MergerRates} we discussed the merger rates of DCBH finding that even under optimistic assumptions, the number of detections that can clearly be assigned to the merger of SMS remnant BHBs is of order unity. We therefore turn the question around and ask what constraining power lies in the non-detection of such sources at $z>15$. Following Poisson statistics, the probability of a non-detection is $p(0)=\exp (-\lambda)$, where $\lambda$ is the expectation value of detections over the observation time $t$, with $\lambda = 0.6 t/\mathrm{yr}$ in our optimistic model. If there is no detection, $1-p(0)$ can be interpreted as the likelihood that this non-detection is representative of the theoretical model. 
Requiring a statistical significance of $95\%$ ($2\sigma$) we derive $\lambda = 3$. Phrased differently, after five years of observations and no detections at $z>15$ we can already exclude our optimistic model with $95\%$ certainty. After 10 years of non-detections, we can rule out the DCBH abundance model that hints at their ubiquity (J$_c=30$) with $3\sigma$ certainty, and set an upper limit of $f_\mathrm{bin}<0.5$ ($0.18$) for $J_c=30$ \J at the $2\sigma$ ($1\sigma$) level.

%%%%%%%%%%%%%%%%%%%%%%%%%%%%%%%%%%%%%%%%%%
\section{Summary}
%%%%%%%%%%%%%%%%%%%%%%%%%%%%%%%%%%%%%%%%%%
Recent studies have been able to derive a range of massive black hole seed number densities, depending on the modelling of realistic physical conditions that lead to their formation.
By choosing two extreme values of J$_c$, the critical LW flux, which is often used to parameterise massive seed formation, we bracket a range of DCBH scenarios ranging from ubiquitous to rare. Recent, high resolution, numerical simulations of the formation of the
first SMBHs in ACHs have suggested that SMSs form in systems
with multiple siblings. By assuming that these systems then go on to host binary black
holes that subsequently merge, we are able to derive an in-situ merger rate per unit redshift for these
black holes. We find the merger rate increases rapidly at high redshift peaking at a redshift of
z $\sim 7$ before declining again as DCBH formation ceases (see Figure \ref{fig:dndtdz}).
  This model can be directly compared to the model of \citetalias{Sesana:2007p2147} for similar
  mass seeds with mergers occuring due to the merger of their host galaxies. While the number
  of mergers is comparable at redshifts around z $\sim 10$, based on optimistic assumptions for the
  binary fraction and coalescence times our model contains significantly more mergers at
  z $\gtrsim 15$ as we only probe the number of in-situ binary mergers in a halo. \\
  \indent We find that if LISA's detector sensitivities can match the design sensitivities then
  massive seeds will be detectable by LISA up to a year before the actual merger with $6.1 \pm$
  2.5 mergers at z $\gtrsim 15$ expected in a ten year mission lifetime. Moreover, if more than
  1 merger is detected at z $\gtrsim 15$ our binary massive black hole model will be verified
  with constraints placed on the number density - $f_\mathrm{bin}$ model (where $f_\mathrm{bin}$ encompassed the
  fraction of heavy seeds that form binaries and merge with short coalescence times).
  Even in the case where no detections
  are made at z $\gtrsim 15$ we will be able to constrain the number density - $f_\mathrm{bin}$ model
  with precision that scales with the mission lifetime. 
  If these statistics can be matched
  with improvements to the IMF, binary fraction, and coalescence times of massive seed black holes
  then strong constraints on the DCBH number density will follow.

%%%%%%%%%%%%%%%%%%%%%%%%%%%%%%%%%%%%%%%%%%%%%%%%%%%%%%%%%%%%%%%%%%%%%%%%%%
%%%%%%%%%%%%%%%%%%%%%%%%%%%%%%%%%%%%%%%%%%%%%%%%%%%%%%%%%%%%%%%%%%%%%%%%%%
  \section*{Acknowledgements}
BA acknowledges the funding from the European Research Council under the European Community's Seventh Framework Programme (FP7/2007-2013) via the ERC Advanced Grant STARLIGHT (project number 339177).  JAR acknowledges the support of the EU Commission through the Marie Sk\l{}odowska-Curie Grant - ``SMARTSTARS" - grant number 699941. TH is a JSPS International Research Fellow. We thank E. Barausse, S. Chon, I. Hosseini, R. Klessen, M. Magg, A. Sesana, and N.Yoshida for useful discussions during the preparation of the manuscript. We thank the referee for helpful comments.
\bibliographystyle{mn2e}
\bibliography{babib}
\label{lastpage}
\end{document}